\documentclass{osa-article}
\journal{oe}

\usepackage{graphicx}
\usepackage{tikz}
\usetikzlibrary{automata,positioning}
\usepackage[T1]{fontenc} 
\usepackage{amsmath}
\graphicspath{{figures/}} 
\usepackage{xcolor}
\usepackage{hyperref}
\hypersetup{urlcolor=blue, colorlinks=true} 
\usepackage{float}
\usepackage{multirow} 
\usepackage{multicol} 
\usepackage{braket}

\begin{document}
\title{Ultrabright Polarization-Entangled Photon Pair Source for Frequency-Multiplexed Quantum Communication in Free-Space}

\author{Emma Brambila\authormark{1,2,*}, Rodrigo G\'omez\authormark{1,2,*}, Riza Fazili\authormark{1,2,$\dagger$}, Markus Gr\"afe\authormark{1,2}, and Fabian Steinlechner\authormark{1,2,$\star$}}
\address{\authormark{1}Fraunhofer Institute for Applied Optics and Precision Engineering IOF,
Albert-Einstein-Straße 7, 07745 Jena, Germany\\
\authormark{2}Institute of Applied Physics, Abbe Center of Photonics, Friedrich Schiller University Jena, Max-Wien-Platz 1, Jena 07743, Germany\\
\authormark{*} both authors contributed equally\\
\authormark{$\dagger$} Current address: INRS-Énergie, Matériaux et Télécommunications, Varennes, Québec, Canada}

\email{\authormark{$\star$}fabian.steinlechner@iof.fraunhofer.de} 

\begin{abstract}
The distribution of entanglement via satellite links will drastically extend the reach of quantum networks. Highly efficient entangled photon sources are an essential requirement towards overcoming high channel loss and achieving practical transmission rates in long-distance satellite downlinks.  Here we report on an ultrabright entangled photon source that is optimized for long-distance free-space transmission. It operates in a wavelength range that is efficiently detected with space-ready single photon avalanche diodes (Si-SPADs), and readily provides pair emission rates that exceed the detector bandwidth (i.e., the temporal resolution). To overcome this limitation, we demultiplex the photon flux into wavelength channels that can be handled by current single photon detector technology. This is achieved efficiently by using the spectral correlations due to hyper-entanglement in polarization and frequency as an auxiliary resource. Combined with recent demonstrations of space-proof source prototypes, these results pave the way to a broadband long-distance entanglement distribution network based on satellites. 
\end{abstract}

\section{Introduction}
Quantum key distribution (QKD) provides distant users (Alice and Bob) with encryption keys whose security can be lower bounded by physics and not unresolved computational complexity arguments~\cite{gisin2002}. Terrestrial QKD systems ~\cite{China2021,EuroQCI,QuNET} are currently limited to metropolitan distances~\cite{China2021} or requires trusted relay nodes~\cite{munro2015} to bridge long distances, and thereby lose any possibility of end-to-end security. End-to-end transmission of quantum states allows secure keys to be generated directly within the secure premises of Alice and Bob. While terrestrial fiber-based QKD has recently been demonstrated up to distances of 200~km in fibers~\cite{Wengerowsky6684} and even 400~km under laboratory conditions~\cite{boaron2018}, the rate decreases exponentially with the fiber length due to attenuation~\cite{NeumannQST}. Extending the reach of quantum links is an ongoing challenge in quantum science and technology, and despite remarkable progress, significant technological and scientific challenges remain on the road to a functional fiber-based quantum repeater architecture~\cite{diamanti_practical_2016,sangouard2011}. %
Optical satellite links are currently the only viable option for long-range transmission of quantum states~\cite{Brito2021}. The recent past has seen rapid technological advances culminate in a series of successful experiments in space~\cite{China2021,Micius2020}. In the context of satellite-based QKD, the entanglement-based approach strongly benefits from the fact that the source in space is not required to be trusted~\cite{gisin2002}. However, due to the substantial attenuation in the dual-downlink scenario, the key rates reported thus far are considered too low for many practical use cases. For example, for the Micius experiment in 2020, the key rate was 0.12 bit/s with an attenuation between 56-71~dB~\cite{Micius2020}. To overcome such limitations, ultrabright and space-qualified entangled photon sources~\cite{EPS-IOF} that are compatible with wavelength multiplexing techniques are central to secure satellite-based quantum key distribution with end-to-end security. 

The prospect of entanglement-based quantum communication has led to remarkable improvement in the performance of polarization entangled sources in the VIS~\cite{EPS_VIS} and especially in the NIR wavelength range~\cite{Anwar}. Here we present the brightest source (as the best of our knowledge) reported to date, whose broad spectral bandwidth makes it an ideal candidate for wavelength demultiplexing. We exploit frequency correlations of the entangled photons to mitigate the increase of accidental pair detection due to higher pump power and limited timing resolution of detectors. We demultiplex the photon flux onto multiple detectors which allows a proportional increase of the possible key-rate, while increasing the average photon pair number directly with the pump power. For a detailed analysis on the optimal working point of entangled photon sources in CW operational regime, we refer to the recent work by Neumann \textit{et al.}~\cite{NeumannPRA}. The improvement via a wavelength demultiplexing scheme to achieve higher channel capacity, and multi-user operation, has been recently reported for the telecom range by Neumann \textit{et al.}~\cite{NeumannArx}. Here we focus on the NIR wavelengths that can be efficiently detected using Si-SPADs. In this wavelength regime there is only one prior experiment, by Pseiner et al.~\cite{Pseiner}, focused on the benefits of demultiplexing for a fixed source emission rate. Here we determine and experimentally demonstrate a source that is capable of reaching the optimal working point (optimal pair rate) of currently available Si-SPAD detection technology.

We realize an ultrabright source of polarization entangled photon pairs and demonstrate its functionality with efficient frequency demultiplexing. Combined with state-of-the-art detectors, this could provide a 136-fold increase of the key rate with respect to the Micius mission\cite{Micius2020}, since our source is able to achieve a rate of 58 bits/s for similar loss conditions for satellite-based QKD. 

Our source, which is based on spontaneous parametric down-conversion (SPDC), generates polarization-entangled photon pairs at a record pair emission rate of 5.5 Mcps per mW of pump power, and 11~nm bandwidth, resulting in potential pair emission rates that are beyond the temporal resolving capacity of state-of-the-art single photon detectors. To overcome the detection limitation, we use the auxiliary spectral correlations that occur naturally as a consequence of energy conservation in the down-conversion process, to demultiplex the photons into pairs of densely spaced frequency channels. Due to the limited availability of off-the-shelf wavelength-division multiplexing components in the wavelength band of space-suitable Si-SPAD detectors, we realize a low-loss demultiplexer based on volume Bragg gratings (VBGs). 

VBGs are an excellent candidate for wavelength demultiplexing in polarization-based quantum communication channels including: high diffraction efficiency (> 99\% achievable~\cite{RizaPW}); low absorption loss~\cite{RizaPW,Moser}; suitability for application in space environment (under radiation, temperature fluctuations and etc.)~\cite{Loicq}; lifespan~\cite{Ott}; wavelength range ($\sim350-2500$~nm) and versatile spectral bandwidth~\cite{Blais}; and  polarization insensitivity~\cite{Elhenny}. Furthermore, we determine the optimal source emission rate for the entire system taking into account the increased quantum bit error rate due to higher order pair emission at high pump powers, as well as the detectors characteristic temporal response function and the electronic coincidence window.


\section{Experimental setup}
\begin{figure}[htb]
	\includegraphics[width=9 cm]{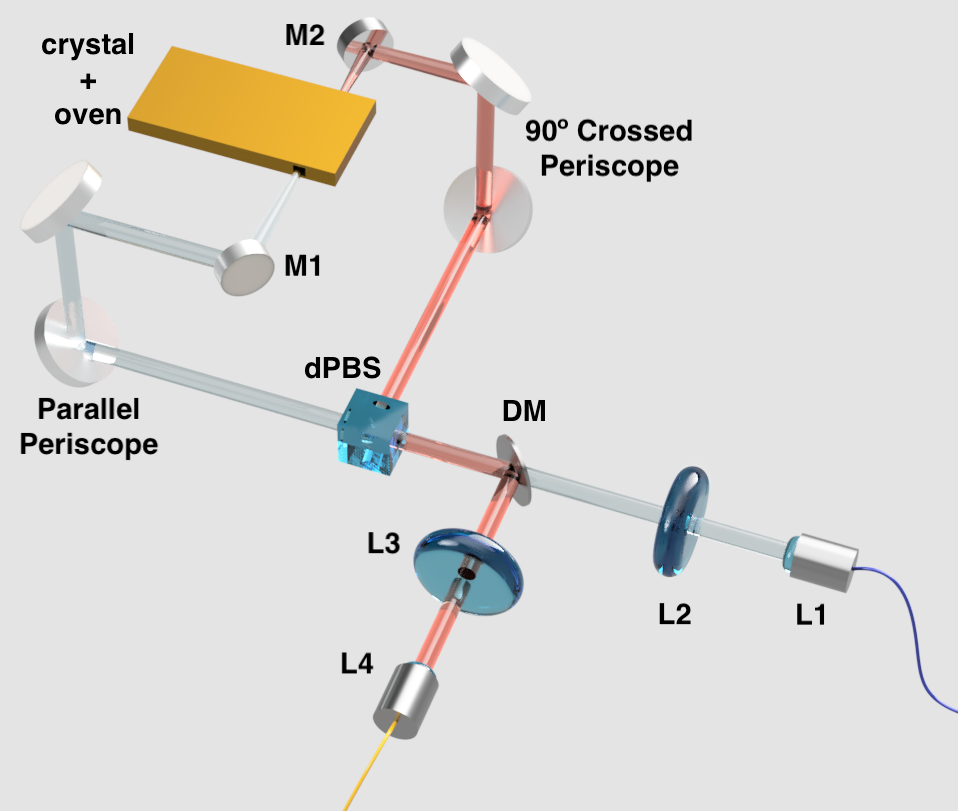}
	\centering
	\caption{Scheme of the entangled photon pair source. Lenses L1 and L2 collimate and focus the pump beam, respectively. A dual-wavelength polarizing beam-splitter (dPBS) separates the diagonally polarized laser pump, and the beam is then steered by mirrors M1 and M2 into the center of the loop. The ppKTP crystal is mounted in a temperature controlled oven at the middle of the loop. A crossed periscope rotates the polarization of the beams, and the parallel periscope preserves the beams' polarization. Finally, lenses L3 and L4 collect signal and idler photons into a single-mode fiber. }
	\label{EPS-0-scheme}
\end{figure}

Our polarization entangled photon source is based on a Sagnac loop configuration~\cite{Kim2006,Fedrizzi2007,Chen2018}, where a periodically poled potassium titanyl phosphate (ppKTP) crystal is bidirectionally pumped by a narrowband laser at 405~nm, such that photon pairs (signal and idler photons) are produced by collinear degenerated type-0 spontaneous parametric down-conversion. A scheme of our experimental source setup is shown in Fig.~\ref{EPS-0-scheme}. There, the pump beam is first collimated and then focused at the geometrical center of the loop with lenses L1 and L2, respectively. A dual-wavelength polarizing beam splitter (dPBS) splits the diagonally polarized pump into its horizontal and vertical polarization components. Then, the incidence of the pump into the ppKTP crystal is adjusted by mirrors M1 and M2, and the crossed periscope rotates the polarization of the beams by 90$^\circ$, while the parallel periscope adjusts the beams' heights without modifying their polarization. The dichroic mirror (DM) separates the down-converted light from the pump beam, transmitting the pump and reflecting the signal and idler light into a common spatial mode. Both signal and idler photons are coupled into a single-mode fiber by lenses L3 and L4. The collection of both photons into a single spatial mode makes the alignment more complex but still allows an operating heralding efficiency, and makes the source compatible with a wide range of wavelength splitting mechanisms. The fiber output is then connected to two different detection scenarios: (a) the output of the single-mode fiber is attached to a 50/50 fiber beam splitter that separates the signal and idler photons with equal probability, and allows to estimate the pair rate of the source; and (b) the output of the single-mode fiber is connected to a wavelength de-multiplexer (DEMUX) that detects signal and idler with certainty by using their frequency correlations.

\begin{figure}[htb]
	\includegraphics[width=12 cm]{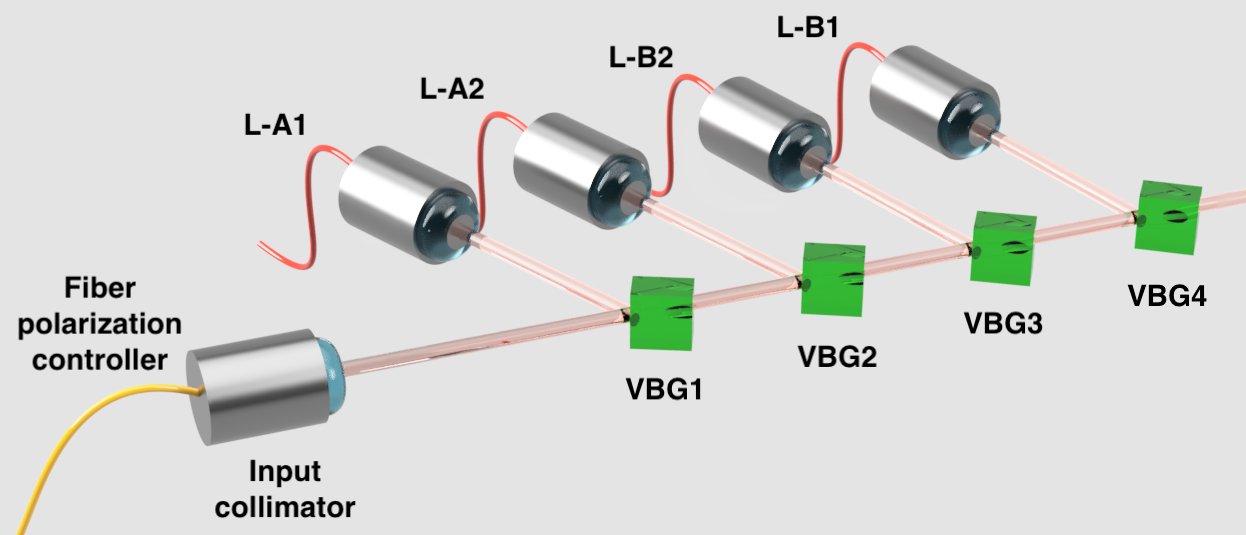}
	\centering
	\caption{Sketch of the low-loss four-channel demultiplexer. The DEMUX uses four volume Bragg gratings (VBG1$\ldots$4), which each reflect a bandwidth of 75 pm from signal and idler photons. The channels are separated by 300~pm to avoid cross-talk between channels.}
	\label{DEMUXFig}
\end{figure}

\begin{figure}[htb]
	\includegraphics[width=12 cm]{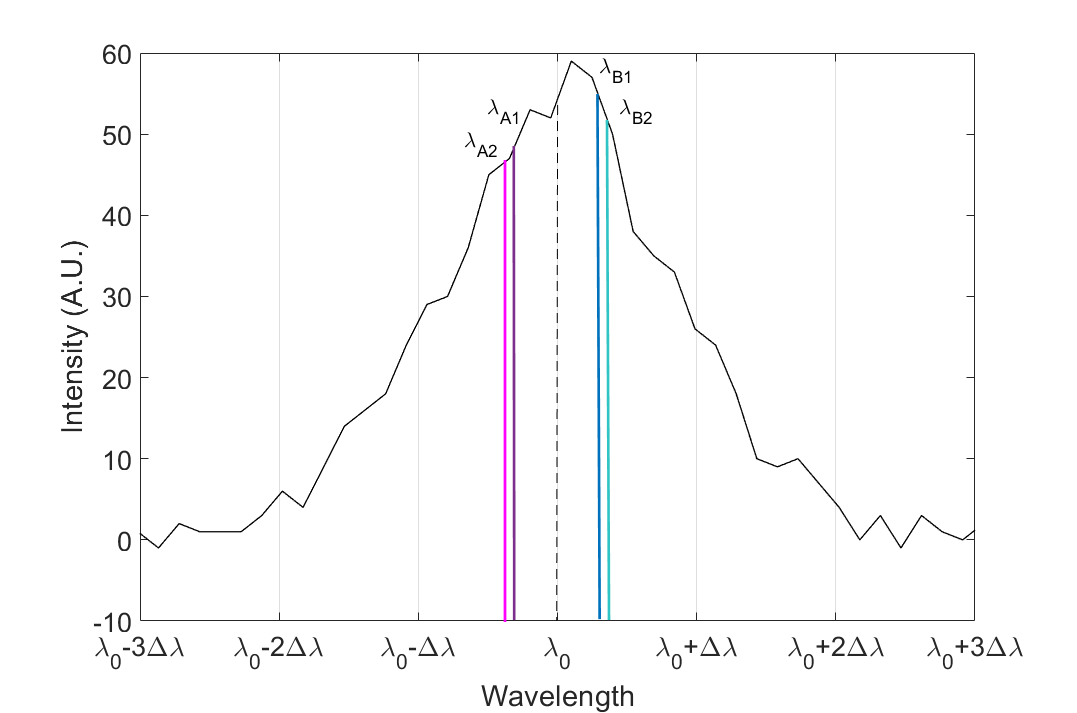}
	\centering
	\caption{Degenerated spectrum of signal and idler photons, taken with a spectrometer with 2~nm resolution. For illustration, here $\Delta\lambda \approx (5 \pm 2)$ nm. The FWHM was characterized to be 11~nm$\pm$ 2~nm. The pump wavelength was measured to be at $\lambda_{\mathrm{p}}= (405.22\pm 0.05)$ nm, with a different spectrometer. The labels $\lambda_{\{\textrm{A1,A2,B1,B2}\}}$ stand for the central wavelengths of the filtered signal photons spectra ($\lambda_{A1,A2}$), and the filtered idler photons spectra ($\lambda_{B1,B2}$) as output of the DEMUX setup, $\lambda_0=2\lambda_{\mathrm{p}}$.}
	\label{SPDC_2}
\end{figure}

The scheme of the DEMUX system is shown in Fig.~\ref{DEMUXFig}. It employs a linear cascade of four photo-thermo-refractive glass reflecting VBGs, each with a diffraction efficiency and spectral bandwidth of approximately 98\% and 75~pm, respectively. Each VBG was tuned to obtain four channels symmetrically around $\lambda_0=810$~nm, with a channel spacing of 300~pm on each side to avoid possible cross-talk (see Fig.~\ref{SPDC_2}). The coupling efficiency towards multimode fibers is approximately 84\%.

Our demultiplexer based on VBGs can distinguish simultaneously four different, resulting in two entangled pairs from our source by using their frequency correlations. Hence, the two-photon state after the DEMUX is:
\begin{align}
    \ket{\psi} \sim& \underbrace{\left[\ket{\lambda_{\text{A1}},\lambda_{\text{B1}}} + \ket{\lambda_{\text{A2}},\lambda_{\text{B2}}} \right]}_{\text{from DEMUX}}\otimes\underbrace{\left[\ket{H_{\text{s}},H_{\text{i}}} + \ket{V_{\text{s}},V_{\text{i}}} \right]}_{\text{from Sagnac loop EPS}}\notag\\
    =&\ket{H_{A1},H_{B1}} + \ket{V_{A1},V_{B1}} + \ket{H_{A2},H_{B2}} + \ket{V_{A2},V_{B2}}\;.
    \label{eq:two_photon_state}
\end{align}
Therefore, a frequency-polarization hyperentangled state is involved. The polarization entanglement  originates from the Sagnac-loop configuration. The frequency part of the state, which stems from energy conservation and the action of the DEMUX, is used as an auxiliary resource. We selected the channels of the demultiplexer in a way that each pair of channels correspond to the wavelength of a signal (s) $\{\lambda_{A1},\lambda_{A2}\}$ and its corresponding idler (i) $\{\lambda_{B1},\lambda_{B2}\}$ photon wavelength, see Fig.~\ref{SPDC_2}.
\section{Experimental Results}
The brightness of the source without DEMUX was measured by connecting the collection single-mode fiber to a 50/50 fiber beam splitter and recording the coincidence counts. From there, the pair rate was extracted to be 5.5 Mcps/mW in a 11~nm bandwidth, and the heralding efficiency was measured to be $>22\%$. This detected rate is found to be the highest compared to previous known experiments, see Tab.~\ref{CPSs}. After correction for losses in the  setup, the emitted brightness at the crystal output is estimated to be >105~Mcps/mW for the full bandwidth. This value was calculated taking into account the transmission through the optics of $\approx 95\%$, fiber coupling of $\approx 50\%$, and the detectors' efficiency of $\approx50\%$, which results in a total detection probability (single photon events) of $\approx 24\%$ at the fiber output of the EPS. 
Utilizing the DEMUX system, we measured a photon pair rate of 80~kcps/mW within the 75~pm bandwidth window, with roughly $20\%$ heralding efficiency.

\begin{table}[htb]
	\begin{centering}
		\begin{tabular}{|c|c|c|c|}
			\hline
			Reference & EPS scheme & Detected pair rate\\
			\hline
			Steinlechner, \textit{et al.} (2014) \cite{FabianPhD} & folded sandwich & $\sim$ 3 Mcps/mW/3nm \\
			Lohrmann, \textit{et al.} (2020) \cite{Lohrmann} & beam displacers interferometer & $\sim$ 120 kcps/mW/0.4nm   \\
			Yin, \textit{et al.} (2020) \cite{Micius2020} & ppKTP in Sagnac loop & 197 kcps/mW/ - \\
			\textbf{EPS of this work} & ppKTP in Sagnac loop & $\sim$ 5.5 Mcps/mW/11nm \\
			\textbf{EPS of this work + DEMUX} & ppKTP in Sagnac loop + VBG &  80 kcps/mW/75 pm\\
			\hline			
		\end{tabular}
		\caption{Summary of recently reported polarization-entangled photon pair sources.}
		\label{CPSs}
	\end{centering}
\end{table}

Even though the wavelength correlations are expected to deterministically split signal and idler photons, we measured some 'cross-talk' coincidences within channels of non-correlated wavelengths (i.e. channels A1-A2, B1-B2, etc). These accidental coincidences are of the same order as the ones expected due to the average single count rate for a fixed jitter detector time, of all channels at a given power. This fact, together with the high coincidence rate between the entangled pairs, implies that the spectral overlap between channels is very low, as well as, the loss of photons pairs at the VBGs, which demonstrates the  optimal functioning of our DEMUX scheme.

\begin{figure}[htb]
\begin{tikzpicture}

\node (img1) {\includegraphics[height=5.3cm]{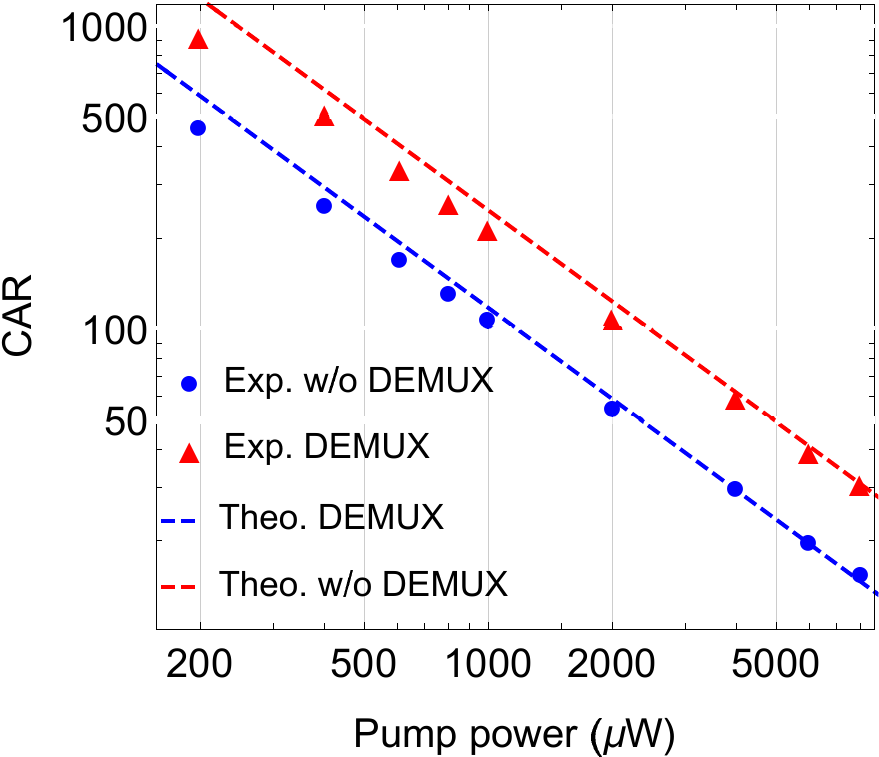}};
\node[left=of img1, node distance=0cm, anchor=center, yshift=2.3cm, xshift=1cm] {{(a)}};

\node[right=of img1,xshift=-0.5cm] (img2) {\includegraphics[height=5.3cm]{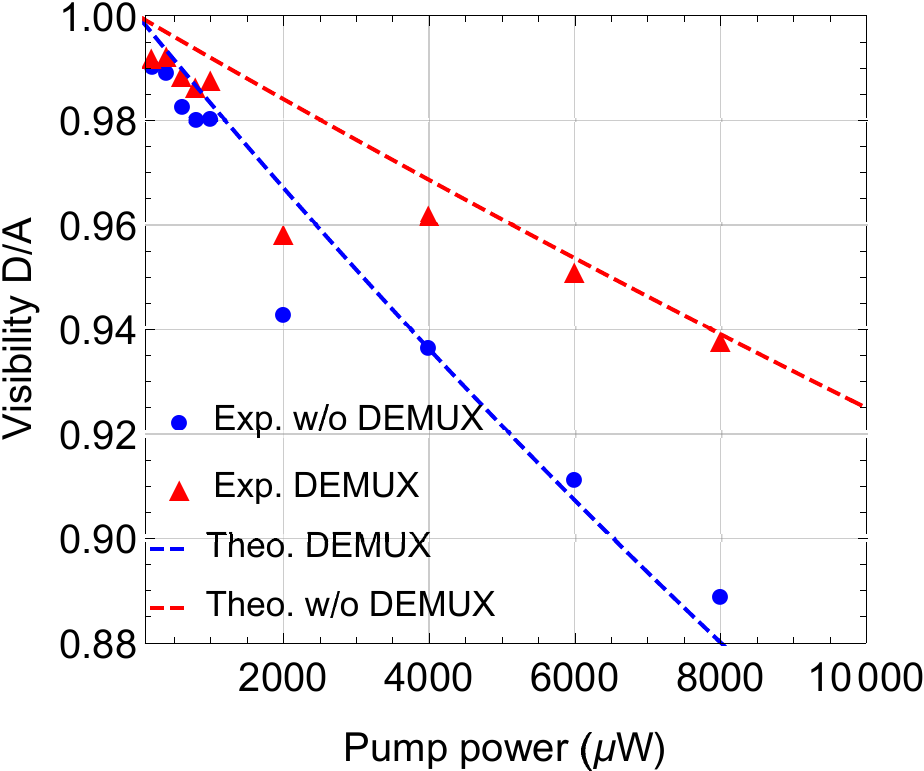}};
\node[left=of img2, node distance=0cm, anchor=center, yshift=2.3cm, xshift=1cm] {{(b)}};
\end{tikzpicture}
\caption{Parameters to quantify the detected entanglement quality, with and without the DEMUX scheme: (a) CAR vs pump power. A clear improvement when for applying the DEMUX is observable. (b) Visibility in the D/A basis for the two entangled pair of photons vs pump power. The use of the DEMUX scheme greatly improves visibility at any pump power.}
\label{fig-car-vis}
\end{figure}

Furthermore, our DEMUX system demonstrates its advantage by measuring the coincidence-to-accidentals ratio (CAR) as it can be seen in Fig.~\ref{fig-car-vis}. The demultiplexer allows a large fraction of accidental coincidences to be filtered (for a large number of wavelength channels they occur primarily between uncorrelated wavelength channels) and results in an improvement on the CAR by $\frac{N}{2}$, where $N$ is the number of channels of the DEMUX and the splitting of each, signal and idler, in two different channels is taken into account. Our experimental results indicate an improvement of the CAR by a factor of two, which agrees with the expectation of using a demultiplexer with four channels. A more detailed discussion is provided in the supplementary information of the article by Lohrmann, \textit{et al.}~\cite{Lohrmann}. 

The photon pairs produced by the Sagnac-loop EPS naturally exhibit a high degree of polarization entanglement (see the two-photon state in Eq.~\ref{eq:two_photon_state}). This is confirmed by visibility measurements in the diagonal/antidiagonal basis, which yield visibility of $\sim99\%$ at low pump power. Fig.~\ref{fig-car-vis} shows the experimentally obtained visibilities in diagonal/antidiagonal basis for several pump powers, with and without the DEMUX. Again, the demultiplexing scheme shows advantages and allows higher visibility for high pump powers. This improvement in the visibility allows to reach higher coincidence rates while maintaining good entanglement quality, which makes our source a suitable candidate for QKD at high rates.

\begin{figure}[htb]
\begin{tikzpicture}
\node (img1) {\includegraphics[height=5.3cm]{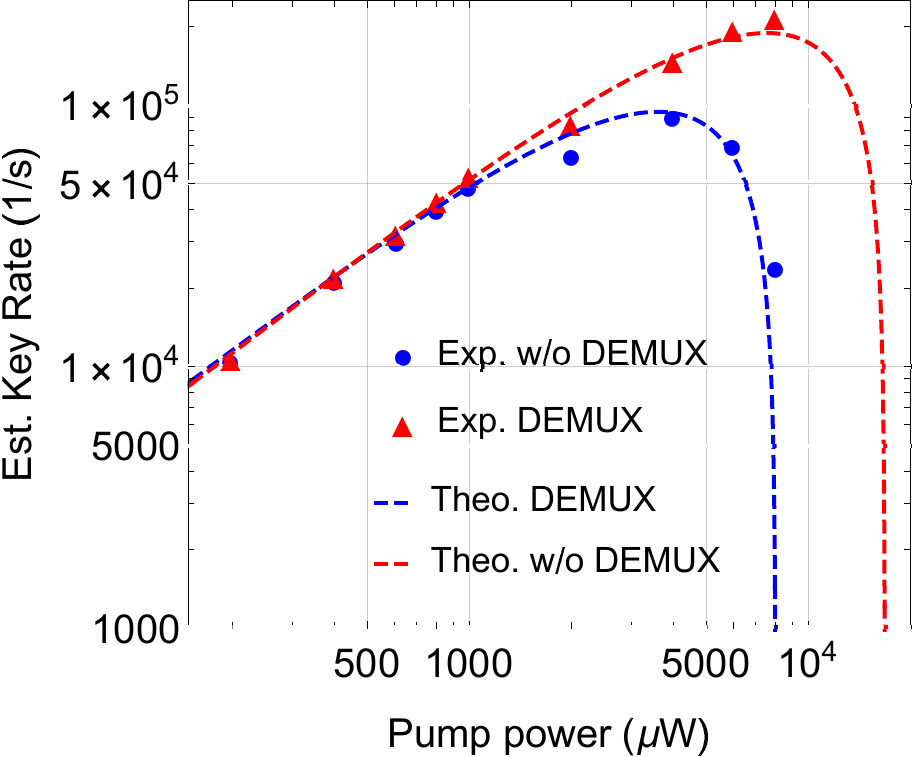}};
\node[left=of img1, node distance=0cm, anchor=center, yshift=2.3cm, xshift=1cm] {{(a)}};
\node[right=of img1,xshift=-0.5cm] (img2) {\includegraphics[height=5.3cm]{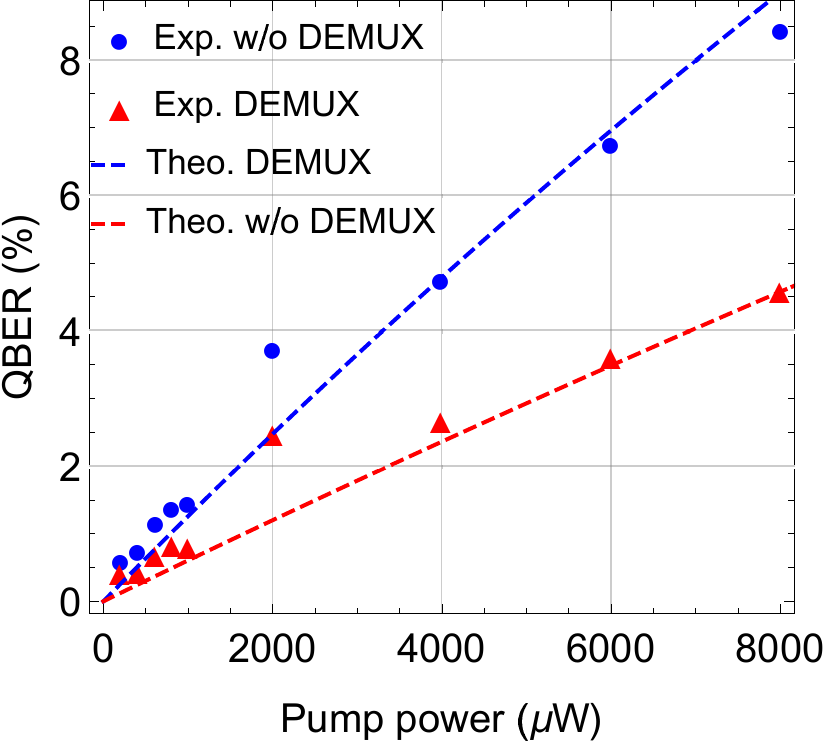}};
\node[left=of img2, node distance=0cm, anchor=center, yshift=2.3cm, xshift=1cm] {{(b)}};
\end{tikzpicture}
\caption{Estimations related to QKD for different average photon pairs as function of the pump power: (a) calculated secure key rate and (b) prediction of quantum bit error rate (QBER), for 3~ns coincidence window. From these graphs, the maximum achievable key rate for the cases of using or not a DEMUX scheme can be compared. Notice that besides the increasing QBER as function of the pump power, the optimization of it due to the demultiplexing allows a much greater maximum key rate. For details refer to the main text.}
\label{fig-skr-qber}
\end{figure}

To also put these results into context with QKD applications over short links, we estimated the back-to-back detection rate, the quantum bit error rate (QBER) and the asymptotic secure key rate for the entanglement based QKD-protocol BBM92~\cite{gisin2002,diamanti2005} with our frequency-multiplexed photon pair source. The results are shown in Fig.~\ref{fig-skr-qber}. One can see that, a strong benefit is that our demultiplexing approach allows higher key rates at high photon pair rate, avoiding high count rate detection issues. In particular,when the coincidence window is 3~ns, we can easily reach with our setup the required mean photon number to obtain the highest key rate possible: with only 9~mW of pump power, it would be possible to achieve a photon detection rate of around 50 Mpairs/s, under laboratory or low loss conditions. The estimation considering a satellite QKD environment by adding higher values of losses can be found at the discussion section.

\begin{figure}[htb]
\begin{tikzpicture}
\node (img1) {\includegraphics[height=5.3cm]{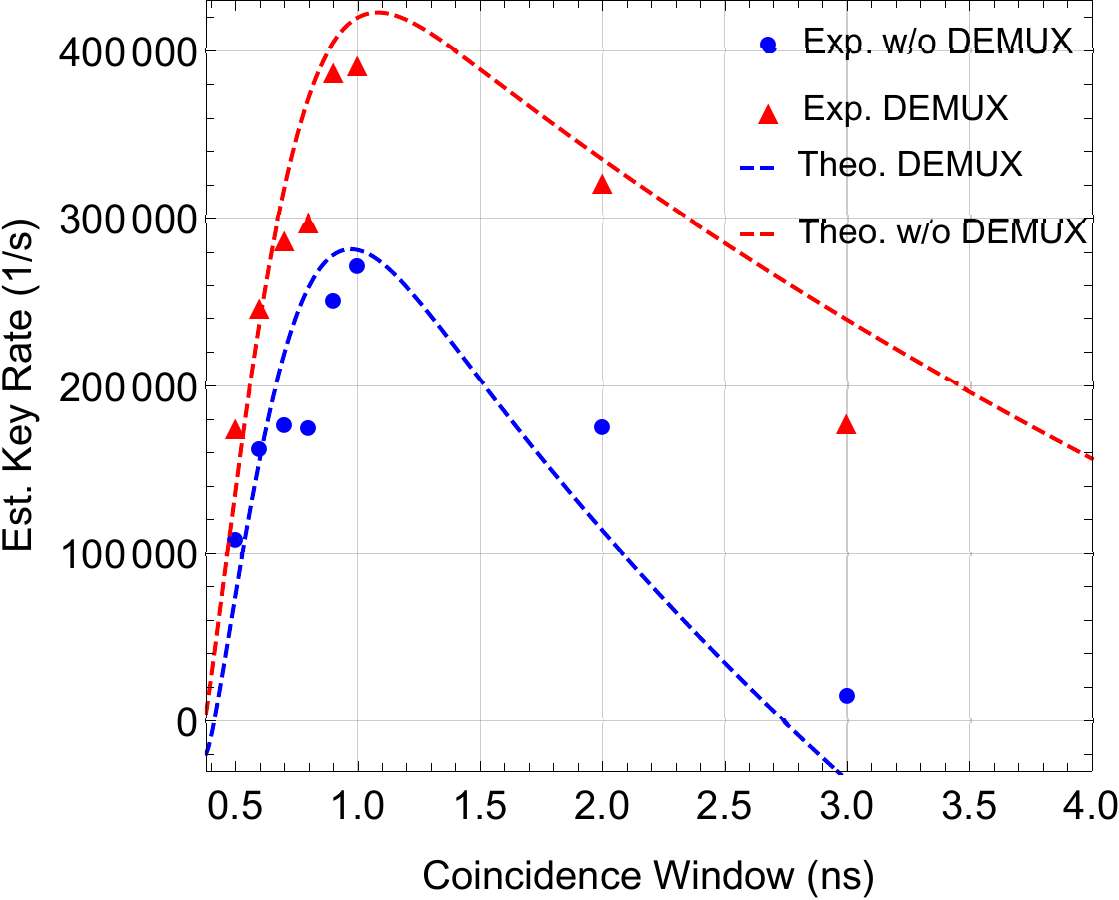}};
\node[left=of img1, node distance=0cm, anchor=center, yshift=2.3cm, xshift=1cm] {{(a)}};
\node[right=of img1,xshift=-0.5cm] (img2) {\includegraphics[height=5.3cm]{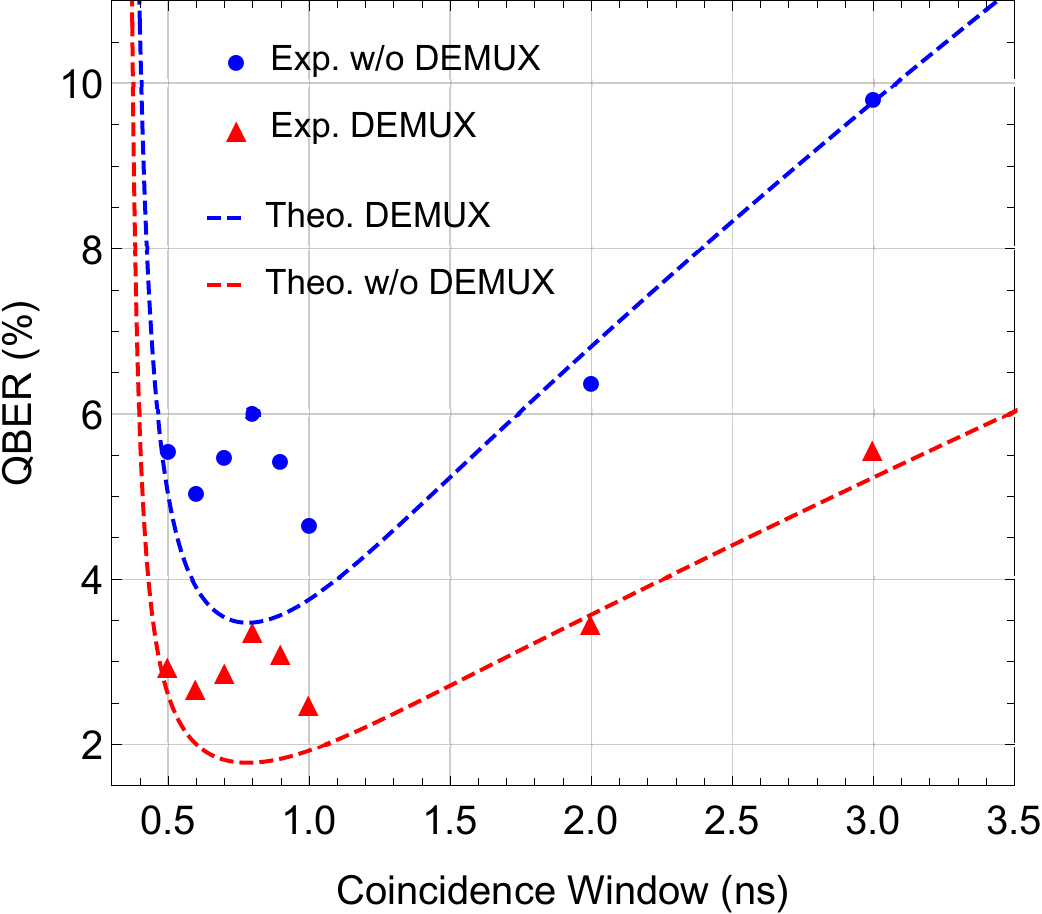}};
\node[left=of img2, node distance=0cm, anchor=center, yshift=2.3cm, xshift=1cm] {{(b)}};
\end{tikzpicture}
\caption{Estimations related to QKD for different average photon pairs as function of the detectors coincidence window: (a) secure key rate and (b) quatum bit error rate (QBER) at 15~mW of pump power. The positive effect of the demultiplexing scheme can be seen in both calculated parameters. Remarkable are the almost doubled key rate and the QBER minima found for a specific coincidence window of approximately 1~ns, which is related to the detectors characteristics. Please refer to the text for further analysis.}
\label{fig-skr-qber-coinc-window}
\end{figure}

\begin{figure}[htb]
\centering
\begin{tikzpicture}
\node (img1) {\includegraphics[height=5.0cm]{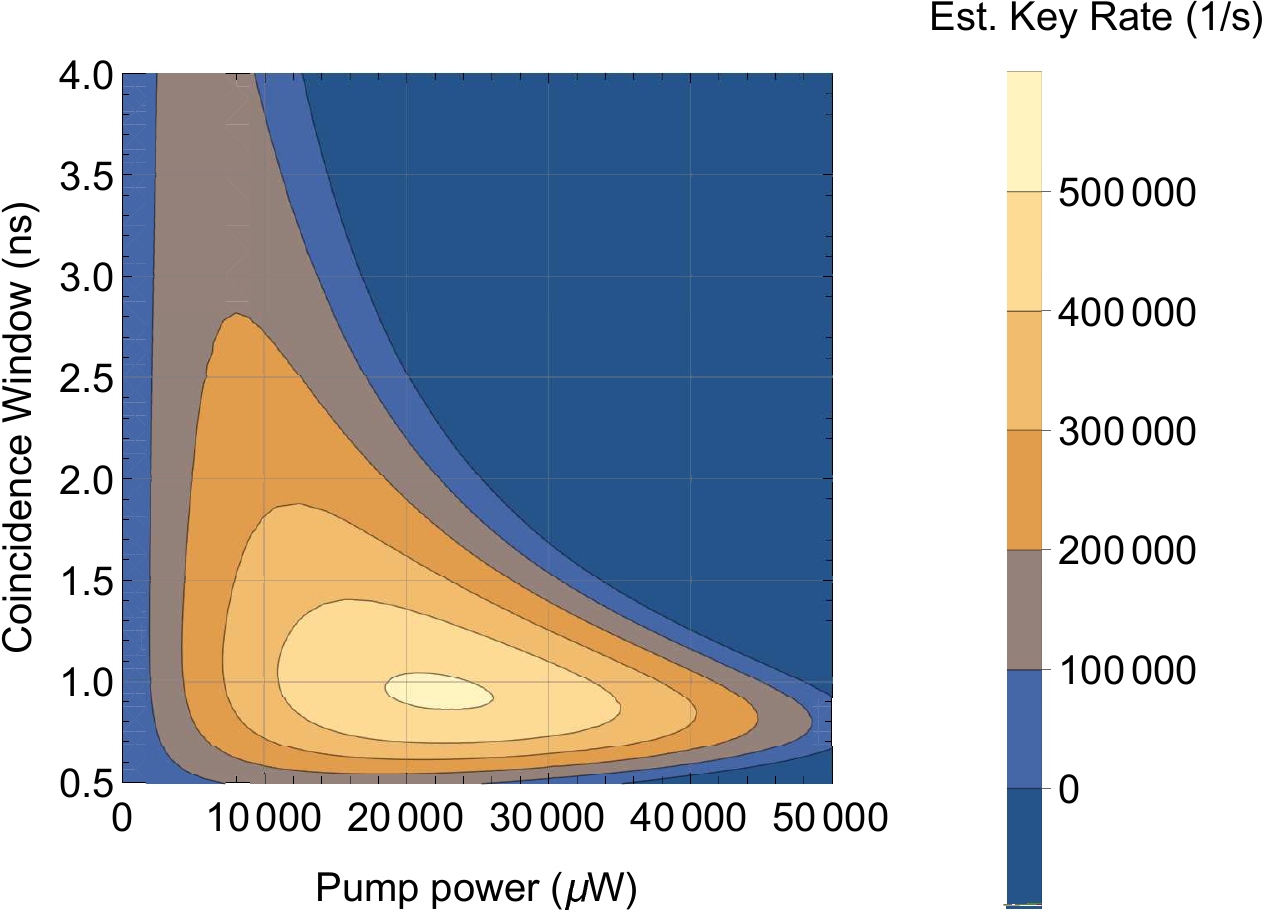}};
\node[left=of img1, node distance=0cm, anchor=center, yshift=2.3cm, xshift=1cm] {{(a)}};
\node[right=of img1,xshift=-0.5cm] (img2) {\includegraphics[height=5.0cm]{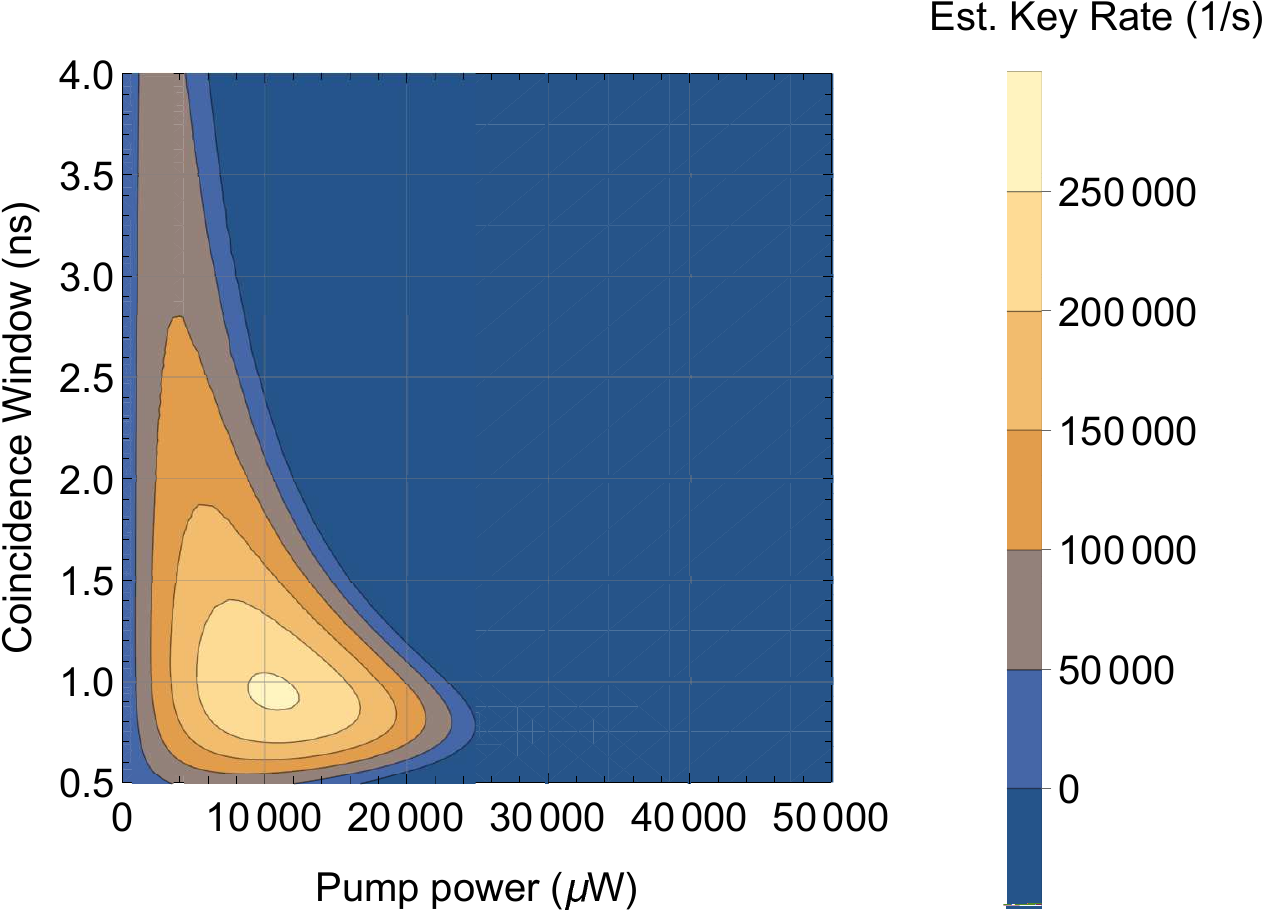}};
\node[left=of img2, node distance=0cm, anchor=center, yshift=2.3cm, xshift=1cm] {{(b)}};
\end{tikzpicture}
\caption{Simulation of the secure key rate in dependency of the pump power and the coincidence window for the cases: (a) with and (b) without using the DEMUX. Clearly, a higher secure key rate becomes possible due to the multiplexing approach, which allows higher pump powers and thus, higher photon pair rates.}
\label{fig-skr-sim}
\end{figure}

By further optimizing the coincidence window, we can increase even more the possible secured key rate that can be obtained with our current ultrabright source, see Fig.~\ref{fig-skr-qber-coinc-window}. As our results show, a coincidence time window in the proximity of 1~ns yields to a maximum rate. This particular value comes from the Si-SPAD detectors characteristics, which normally have a time resolution of $\approx$ 0.4~ns. In the same manner, one can minimize the QBER. Notice that its minimum locates slightly lower than 1~ns, however its increase at exactly 1ns is insignificant when the maximum key rate is estimated. 
Following the inspection of the effect due to the pump power and coincidence window as experimental paramenters, we simulated the combined dependency of the secure key rate with and without the DEMUX with our current experimental capabilities, see Fig.~\ref{fig-skr-sim}. While the optimal coincidence window stays in the proximity of 1~ns, the optimal pump power increases by a factor of two in the case of the DEMUX, resulting in a doubled secure key rate. To improve even further the possible key rate, one would need to increase the heralding of the source, increase the detection efficiency, and reduce the coincidence window by having detectors with higher time resolution. For example, by employing state-of-the-art superconducting nanowire single photon detectors, with a typical coincidence window of the order of 15 ps and an detection efficiency above 90\% at $\sim810$~nm, the accidental coincidences would significantly reduce and one could take even more advantage of even higher pump powers and correspondingly increased photon pair flux. 

\section{Discussion}
\begin{figure}[htb]
\centering
\includegraphics[height=6.5cm]{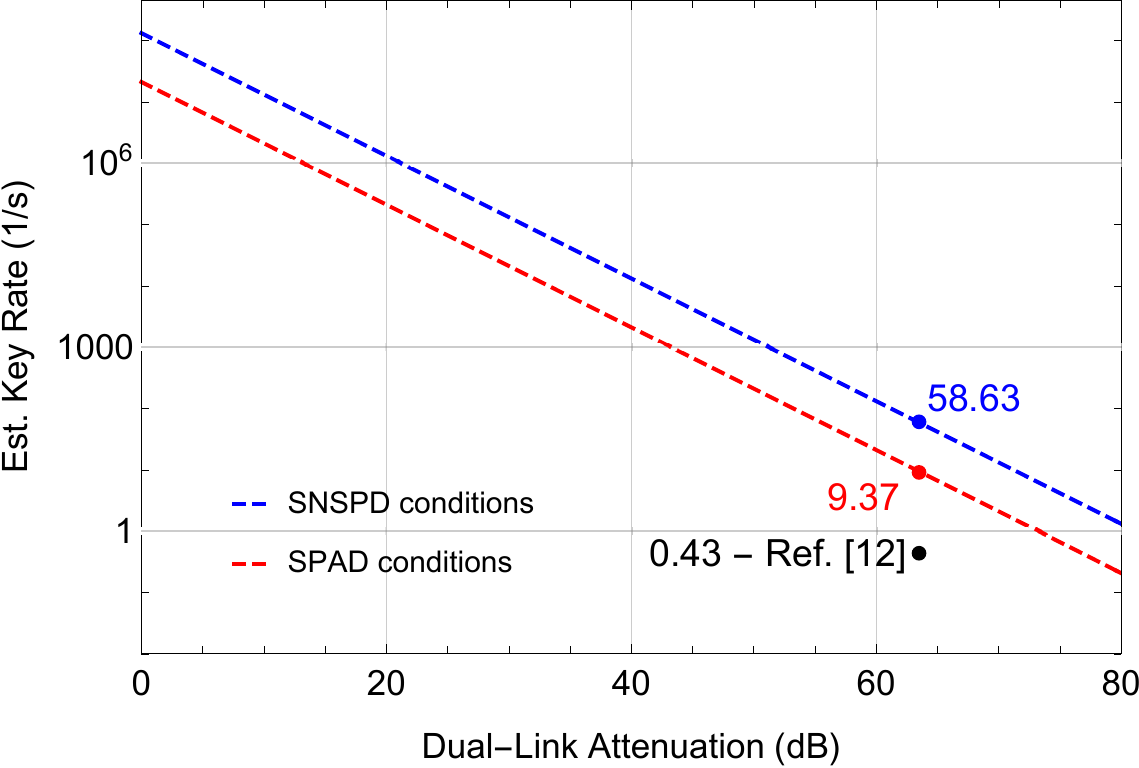}
\caption{Estimated secure key rate (asymptotic limit) for dual-down link QKD scenario as a function of the dual-link attenuation assuming symmetric losses for both photons. On each detection side, 56 detection channels are assumed. For the SNSPD conditions, we consider the detectors' efficiency of 80\%, coincidence window of 100~ps, and 100~Hz detector dark counts.For the SPAD conditions, we consider the detector efficiency of 50\%, coincidence window of 2.5ns and 100 Hz dark counts. The black dot correspond to the asymptoptic rate of the MICIUS mission~\cite{Micius2020}. }
\label{fig:dual_down_link}
\end{figure}

The characteristic broad spectral bandwidth of our EPS, which can be seen in Fig.~\ref{SPDC_2}, would allow us to de-multiplex approximately 56 channels by using VBG of 75~pm of FWHM. Hence, our EPS is capable of supporting 28 QKD channels at once. On the one side, this renders multiuser QKD scenarios more feasible. On the other side, two-party QKD scenarios between Alice and Bob will benefit from an extreme boost in the secure key rate by applying the multiplexing scheme. To elaborate on the advantages of the presented multiplexing scheme together with our ultrabright EPS, we evaluate the performance boost that it would bring to long-distance satellite-based QKD. Assuming similar conditions as the MICIUS mission~\cite{Micius2020}, and using current detection technology based on superconductor nanowire single photon detectors, it would be possible to achieve key rates of the order of 58~bits/s for a 63~dB dual-link with symmetric losses. Moreover, this comes along with a 28-fold improved CAR. The scaling of the secure key rate over different dual-link attenuation, for different coincidence windows, is plotted in Fig.~\ref{fig:dual_down_link}. 

This demonstrates that QKD at high rates becomes feasible by harnessing ultrabright polarization entangled photon pair sources, and frequency multiplexed detection on Alice's and Bob's side. Therefore, the ultrabright EPS presented here together with our free-space DEMUX is a good candidate to improve satellite-based QKD and thus, efficiently enable long-distance secure quantum communication links.

Recently, Lohrmann \textit{et al.}~\cite{Lohrmann} reported a source that generates high-fidelity polarization entanglement over a spectral range of 100~nm. In combination with low-loss frequency multiplexing, such a configuration provides a clear path towards achieving even higher total transmission rates using state-of-the-art detection technology on ground. The results thus represent an important step towards the feasibility of recently proposed experiments for space-borne entanglement-based quantum key distribution via high-loss channels, such as dual down-links from geostationary satellites.

\section{Summary}
In our work, we experimentally demonstrate a frequency multiplexing approach for polarization entangled photon pairs to overcome the limited timing resolution of current time-resolved single-photon detection systems. To this end, we have realized an ultra-bright source of polarization entangled photon pairs and a low-loss demultiplexer. The source is based on a Sagnac loop configuration and provides  (to the best of our knowledge) unmatched performance with respect to brightness in combination with visibility - see~\cite{Anwar} for a review of performance characteristics of sources to date. By spectrally multiplexing on signal and idler side, we are able to demonstrate an increased visibility, coincidence to accidental ratio and finally an improved possible quantum bit error rate and secure key rate. We extent our experimental findings with simulations on possible satellite-base free space links and show a tremendous improvement in the achievable secure key rate applying our ultrabright source with the multiplexing scheme and latest photon detection systems. Our work paves the way to more suitable long-distance QKD scenario with exploiting state of the art technology.

\section*{Funding}
This work was supported by the Fraunhofer LIGHTHOUSE PROJECT (QUILT) and by the Fraunhofer Attract program (QCtech). The presented results were partially acquired using devices funded by the Federal Ministry of Education and Research of Germany (BMBF) through the project Quantum Photonics Labs (QPL) with the funding ID 13N15088.


\section*{Acknowledgments}
Tha authors acknowledge financial support from the Fraunhofer Society and the Federal Ministry of Education and Research of Germany (BMBF). 

\section*{Disclosures}

The authors declare no conflicts of interest.

\section*{Data Availability}

The data that support the findings of this study are available from the corresponding authors upon reasonable request.

\bibliography{Bibliography}

\end{document}